\documentclass[aps,showpacs,prl,manuscript]{revtex4}

\usepackage{amsmath}
\usepackage{amssymb}
\usepackage{amscd}
\usepackage{wasysym}
\usepackage[ansinew]{inputenc}
\usepackage[T1]{fontenc}
\usepackage{ae,aecompl}

    \usepackage[dvips]{graphicx}
    \DeclareGraphicsExtensions{.eps}

\newcommand{\be}{\begin{equation}}
\newcommand{\ee}{\end{equation}}
\newcommand{\bea}{\begin{eqnarray}}
\newcommand{\eea}{\end{eqnarray}}
\newcommand{\nn}{\nonumber}

\newcommand{\Ps}{\mathcal{P}}
\newcommand{\HH}{\mathcal{H}}

\begin{document}
\bibliographystyle{apsrev}

\title{Kuramoto dynamics in Hamiltonian systems}

\author{Dirk Witthaut$^1$} \author{Marc Timme$^{1,2}$}
\affiliation{$^1$ Network Dynamics Group, Max Planck Institute for Dynamics and Self-Organization (MPIDS),
37077 G\"ottingen, Germany}
\affiliation{$^2$ Faculty of Physics, University of G\"ottingen, 37077 G\"ottingen, Germany}

\date{\today }

\begin{abstract}
The Kuramoto model constitutes a paradigmatic model for the
dissipative collective dynamics of coupled oscillators, characterizing in
particular the emergence of synchrony. Here we present a classical Hamiltonian (and thus conservative)
 system with $2N$ state variables that in its action-angle representation
exactly yields Kuramoto dynamics on
$N$-dimensional invariant manifolds. We show that the synchronization
transition on a Kuramoto manifold emerges where the transverse
Hamiltonian action dynamics becomes unstable. The uncovered Kuramoto dynamics in Hamiltonian systems thus distinctly links dissipative to conservative dynamics.
\end{abstract}

\pacs{05.45.Xt}

\maketitle
Spontaneous synchronization
constitutes one of the most prevalent order forming processes in Nature \cite{Piko03}.
In 1975, Kuramoto introduced a now standard model of weakly coupled limit
cycle oscillators to analyze synchronization processes \cite{Kura75}. The model characterizes the collective dynamics of a variety of 
dynamical systems ranging from
chemical reactions  \cite{Kura84}
and  neural networks \cite{Somp90}
to coupled Josephson junctions \cite{Wies96},  laser arrays \cite{Vlad03}
and optomechanical systems \cite{Hein11}. 
In the Kuramoto model, $N$ phase
oscillators are coupled via their phase differences. The rate of change of each phase $\phi_j$ is given by 
\be
  \frac{d \phi_j}{dt} = \omega_j + \frac{K}{N}
     \sum_{\ell = 1}^{N} \sin(\phi_\ell - \phi_j),
     \label{eqn:kuramoto-intro}
\ee
where $\omega_j$ is the intrinsic frequency of the $j$th oscillator, $j \in \{1,\ldots,N\}$,  and $K$ 
denotes the coupling strength. Often, the frequencies are randomly drawn from a distribution $g(\omega)$ with finite width.
If $K$ exceeds a certain threshold $K_c$, this
system exhibits a phase transition from an incoherent to a synchronous, phase-ordered asymptotic state in the thermodynamic limit $N\rightarrow \infty$.

Despite its broad importance, many features of the Kuramoto model remain
unknown. In particular, several of its relaxation and stability 
properties and the collective dynamics for finitely many coupled oscillators
seem unusual for a dissipative system and are still not fully understood \cite{Stro00, Aceb05, Mais04,Kori04, Grab10, Mais04}.

In this Letter, we introduce a class of (classical) Hamiltonian systems that exhibit a family of invariant tori on which the
dynamics is identical to that of the Kuramoto model
(\ref{eqn:kuramoto-intro}). After demonstrating mathematical equivalence of these dynamics, we numerically and analytically study the full
volume-preserving Hamiltonian dynamics, focusing on the synchronization
transition and its consequences. Intriguingly, the onset of
synchronization implies the onset of transverse instability of the invariant tori. We derive an analytic expression quantifying the (phase) order parameter in terms of the local action instability. Beyond local dynamics, the deviation from the tori measured by the inverse participation ratio of the Hamiltonian system provides a distinguished indicator for the synchronization transition. It even scales more favorably with system size than the standard synchronization order parameter. Studying the proposed Hamiltonian
systems may thus help to better understand the collective dynamics of the Kuramoto model, in particular for finite $N$. 

\begin{figure*}[bt]
\centering
\includegraphics[width=5cm, angle=0]{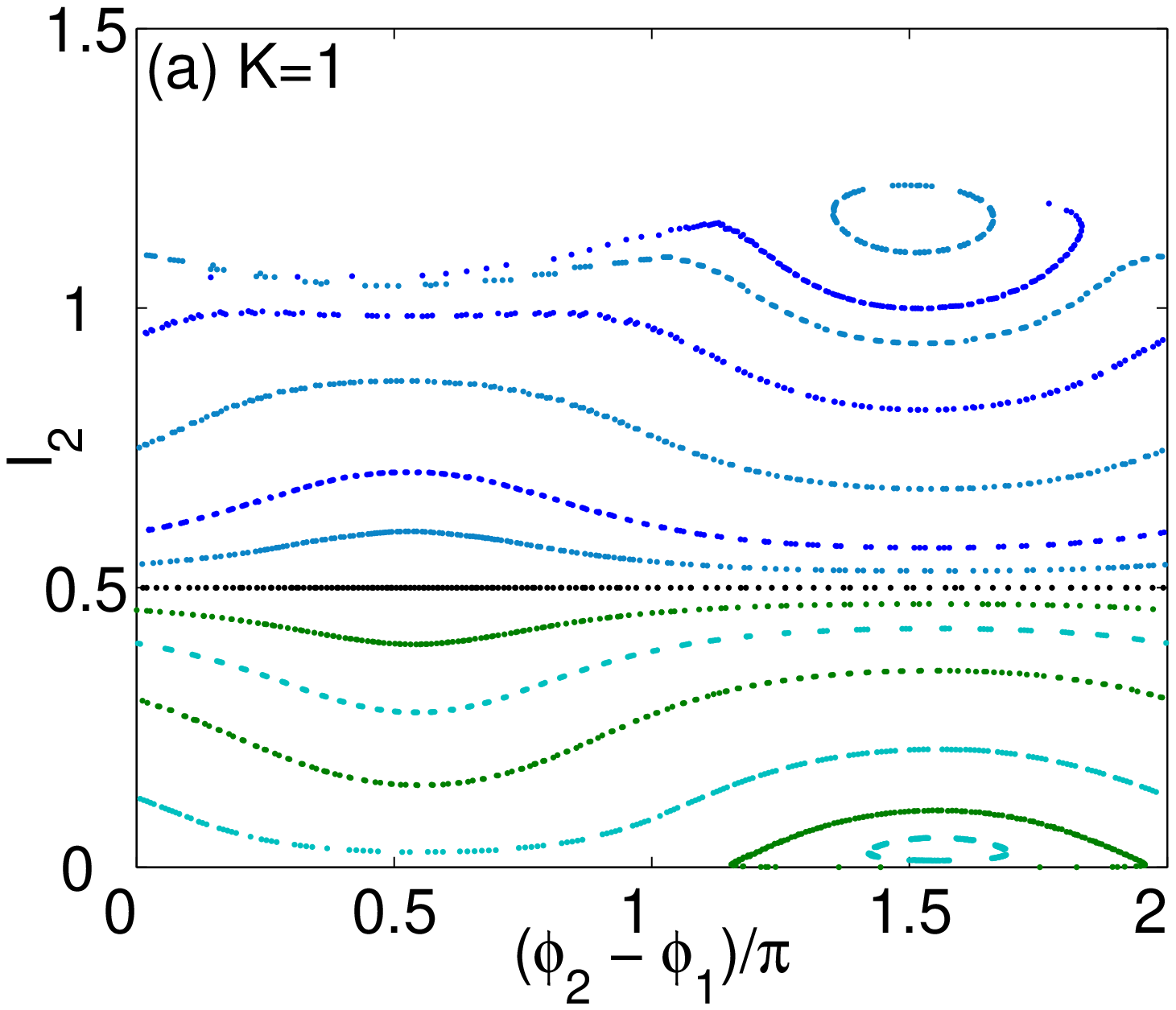}
\includegraphics[width=5cm, angle=0]{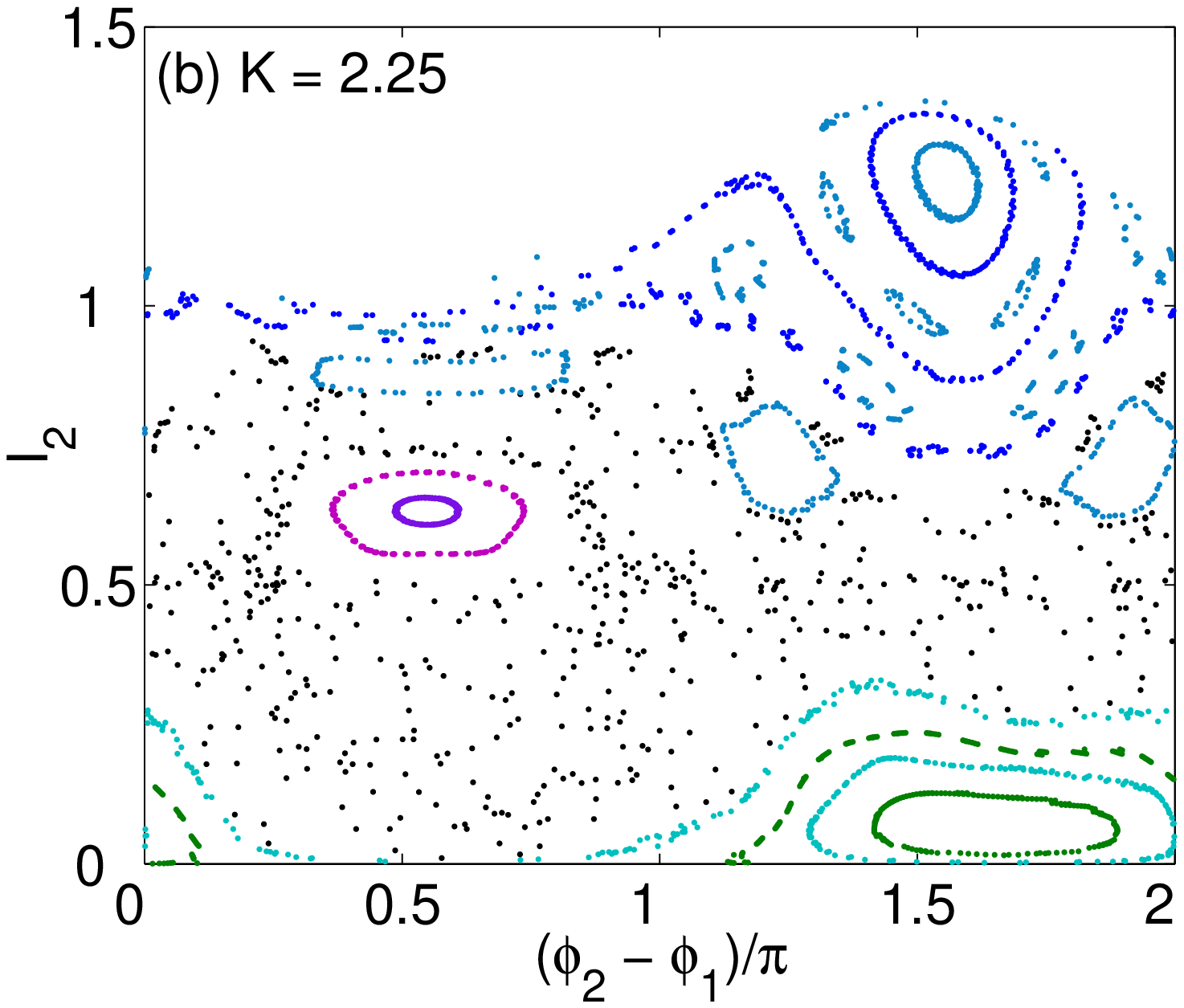}
\includegraphics[width=5cm, angle=0]{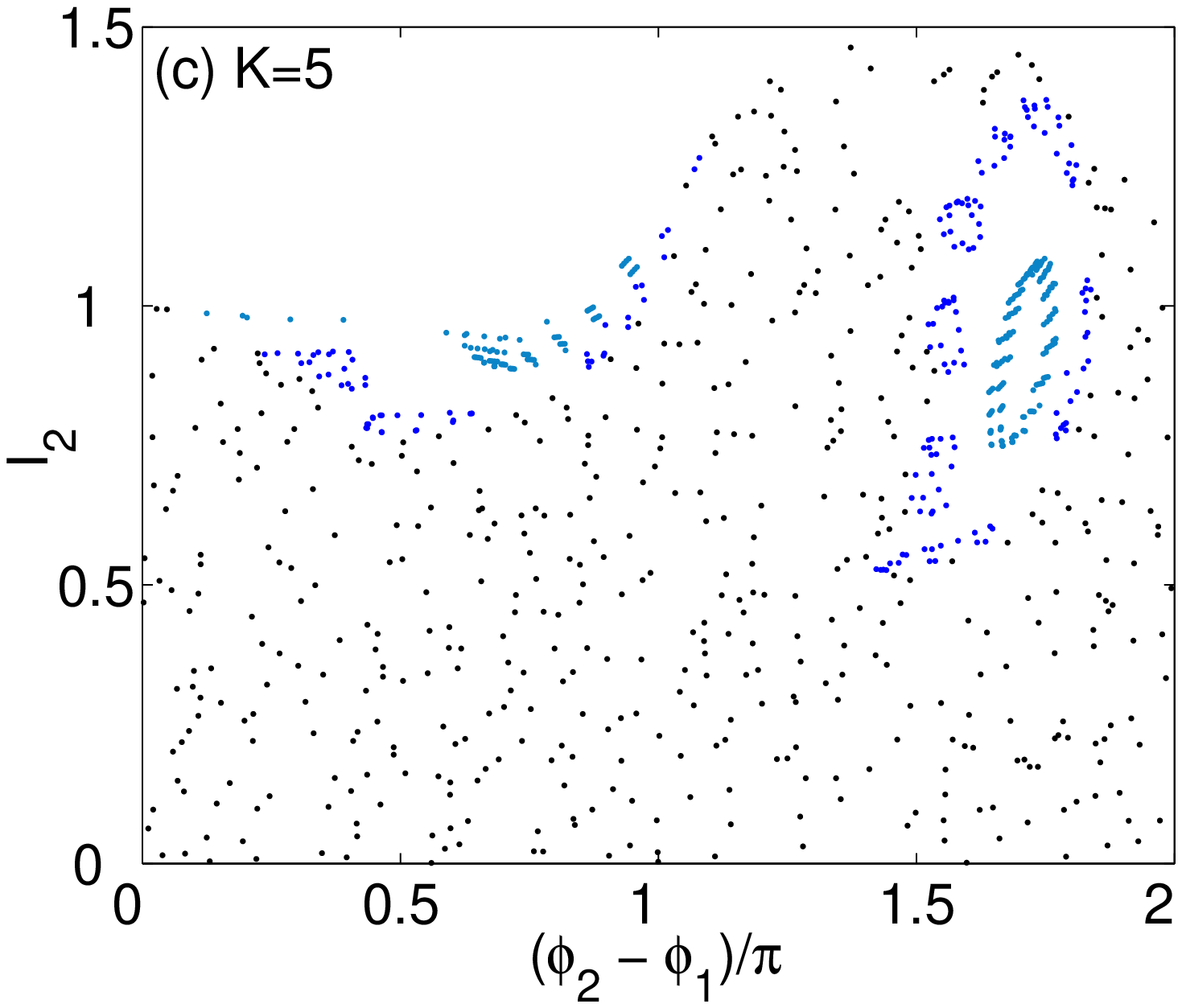}
\caption{\label{fig:phase}
Emergence of chaos in the full Hamiltonian dynamics?
Panels display two-dimensional Poincar\'e section of the phase space  for $N =
3$, where the six-dimensional phase space reduces to effectively $2N-3=3$ 
dimensions \cite{psection}.
Different colors are used to guide the eye.
The oscillator frequencies are $(\omega_1,\omega_2,\omega_3) = (-2,-1,3)$.
(a) $K=1<K_{c1}$, (b) $K_{c1}<K=2.25<K_{c2}$, (c) $K=5>K_{c2}$.
}
\end{figure*}

Consider the Hamiltonian function
\bea
  \label{eqn:ham1}
  && \HH'(q_1, p_1, \ldots, q_N, p_N) 
      = \sum_{\ell = 1}^N \frac{\omega_\ell}{2} (q_\ell^2 +  p_\ell^2)  \\
      && \qquad + \frac{K_0}{4N} \sum_{\ell,m = 1}^{N}
       (q_\ell p_m - q_m p_\ell) ( q_m^2 + p_m^2 - q_\ell^2 - p_\ell^2 ) \nn  
\eea 
defined on the $N$-particle phase space $\mathbb{R}^{2N}$. 
The canonical transformation
\be
  I_\ell =  \left( q_\ell^2 + p_\ell^2 \right)/2 \quad
 \mbox{and} \quad
  \phi_\ell = \arctan \left( q_\ell/p_\ell \right)
\ee
for  $\ell \in \{1,\ldots,N\}$  simplifies the representation of the Hamiltonian in terms of action-angle variables $I_\ell$ and $\phi_\ell$ for uncoupled ($K_0=0$) harmonic oscillators (with
single-particle Hamiltonian $\HH'_\ell(q_\ell,p_\ell)=\omega_\ell \left(q_\ell^2 + p_\ell^2
\right)/2$). This transformation is invertible if and only if all $I_\ell > 0$, in particular if all  $\HH'_\ell>0$.
In the new action-angle variables $(\mathbf{I},\boldsymbol{\phi}) \in
\mathbb{R}^N_+ \times \mathbb{S}^N$, the Hamiltonian reads
\bea
  \label{eqn:ham1_iphi}
  && \HH(I_1, \phi_1, \ldots, I_N, \phi_N) 
      = \sum_{\ell = 1}^N \omega_\ell I_\ell  \\
      && \qquad - \frac{K_0}{N} \sum_{\ell,m = 1}^{N} 
              \sqrt{I_m I_\ell} (I_m  - I_\ell) \sin(\phi_m - \phi_\ell) \nn
\eea 
and the equations of motion are given by
\bea
     \label{eqn:eom_iphi}
    \dot I_j &=& - \frac{\partial \HH}{\partial \phi_j}  \\
   &=& - \frac{2 K_0}{N} \sum_{m=1}^N \sqrt{I_m I_j} (I_m - I_j) 
                 \cos(\phi_m - \phi_j), \nn \\
    \label{eqn:eom_iphi2}
  \dot \phi_j &=& \frac{\partial \HH}{\partial I_j}
  = \omega_j + \frac{K_0}{N}   
        \sum_{m=1}^N \left[ 2 \sqrt{I_j I_m} \sin(\phi_m - \phi_j)  \right.\\
      && \qquad \qquad \qquad   \left. +
          \sqrt{I_m/I_j} (I_m - I_j) \sin(\phi_m - \phi_j)  \nn \right] .
\eea
One important property of these equations of motion is that
they leave specific manifolds invariant. Any state with all actions homogeneous, $I_j \equiv I > 0$ for all $j$, yields $d I_j/dt = 0$ and thus leaves all actions unchanged. Thus the family of toric manifolds
\be
  \label{eqn:deftori}
   T^{N}_{I}  = \left\{ (\mathbf{I},\boldsymbol{\phi}) \in 
    \mathbb{R}^N_+ \times \mathbb{S}^N \ | \  
       \forall j \in \{1,\ldots, N \} :  I_j = I \right\} 
\ee
are invariant under the flow generated by (\ref{eqn:eom_iphi}) and (\ref{eqn:eom_iphi2}) for each given $I$. On one given torus, i.e.~for one value of $I$, the dynamics of the phases
\be
   \dot \phi_j = \omega_j + \frac{2 I K_0}{N}
     \sum_{\ell = 1}^N \sin(\phi_\ell - \phi_j), 
     \label{eqn:kuramoto-red1}
\ee
equals that of the original Kuramoto model (\ref{eqn:kuramoto-intro})
with a coupling constant $K=2 I K_0$.
We conclude that the Hamiltonian function (\ref{eqn:ham1})
generates the Kuramoto model  on the invariant tori $T_I$. This holds for all coupling  strengths and arbitrary frequency distributions.

The Hamiltonian dynamical system defined by 
(\ref{eqn:eom_iphi}), (\ref{eqn:eom_iphi2}) has two constants of motion, the
Hamiltonian function $\HH$ itself and (twice) the total action
\bea
   C^2 &=& \sum \nolimits_{j=1}^N (p_j^2 + q_j^2) 
   = 2 \sum \nolimits_{j=1}^N I_j.
             \label{eqn:constantom}
\eea
The dynamics (\ref{eqn:eom_iphi}) is equivariant under a simultaneous 
scaling transformation $(p_j,q_j) \rightarrow  (Cq_j, Cp_j)$ for all $j$
and $K \rightarrow K/C^2$ for every $C>0$, so we fix the normalization as 
$
C^2=N
$
such that $K=K_0$ without loss of generality and $I=1/2$ defines the Kuramoto manifold $T^{N}_{1/2}$. Furthermore, the dynamics is equivariant with respect to a global phase shift, because it depends only on the phase differences. The two constants of motion and the shift-equivariance make the state space of the full Hamiltonian system effectively $(2N-3)$-dimensional, while the invariant subspace $T_{1/2}^N$ is an $(N-1)$-dimensional torus. In the following we drop the subscript
$1/2$ for convenience.

What does the Hamiltonian dynamics tell us about the Kuramoto 
dynamics on the invariant manifold $T^N$?  For the simplest system 
of $N=2$ units \cite{Witt12} there are two elliptic fixed points off 
the Kuramoto manifold below a critical coupling $K<K_c=|\omega_1-\omega_2|$;  
when $K$ becomes larger than $K_c$, two additional elliptic fixed points emerge 
off and two hyperbolic ('saddle') points on $T^{N}$ via a Hamiltonian saddle 
node bifurcation. One of the hyperbolic fixed points implies phase locking 
as it is stable within  $T^N$ and unstable transverse to it.

Systems with $N \ge 3$ show a much richer dynamics and suggest the 
emergence of chaos (cf.~Fig.~\ref{fig:phase}). 
Consider for instance three units with frequencies 
$(\omega_1,\omega_2,\omega_3) = (-2,-1,3)$.
Whereas for small coupling, $K<K_{c1}\approx 1.5$, the Poincare section  \cite{psection} indicates exclusively regular Hamiltonian dynamics (Fig.~\ref{fig:phase}a), irregular dynamics (Fig.~\ref{fig:phase}c) prevails for sufficiently large coupling,  $K>K_{c2}\approx 4.6$, with mixed state space  (Fig.~\ref{fig:phase}b) for intermediate $K$. Simultaneous to the transitions in the full Hamiltonian system, synchronization appears on the Kuramoto manifold $T^N$ as
$K$ increases. The oscillators 1 and 2 are unlocked for weak coupling and phase lock for $K > K_{c1}$,  while the phase of the third oscillator remains incoherent.  For $K > K_{c2}$, 
global phase locking sets in. For all $K\in[0,\infty)$, the dynamics 
is non-chaotic within the Kuramoto manifold (compare also to \cite{Wata93,Ott08,Bick11}).

\begin{figure}[tb]
\centering
\includegraphics[width=8cm, angle=0]{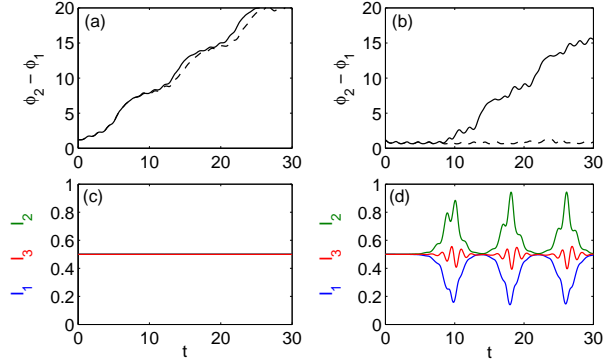}
\caption{\label{fig:3osc2}
Action bursts indicate onset of synchrony. Panels show the dynamics of three coupled oscillators in the regime of
(a,c) no phase locking and (b,d) partial phase locking.
Solid lines show the dynamics of (a,b) the phases and (c,d)
the actions $I_j(t)$. The initial state is drawn randomly by perturbing the actions $I_j(0)$ off the manifold $I_j\equiv 1/2$ by a random amount of the order of $10^{-4}$; it is thus close to 
but not on $T^N$.
Dashed lines show the dynamics of the phases $\phi_j(t)$
on the submanifold $T^N$, i.e.~Kuramoto dynamics. 
Parameters are $(\omega_1,\omega_2,\omega_3) = (-2,-1,3)$ and 
(a,c) $K=1$ and (b,d) $K = 2.25$, respectively.
}
\end{figure}

Phase-locking is indeed closely linked to the instability of the Hamiltonian action dynamics:
For small coupling, where the Kuramoto dynamics is not phase-locked, the actions 
exhibit stable dynamics  (cf.~Fig.~\ref{fig:3osc2} (a) and (c)) and 
 ``action locking''  $I_1 \approx I_2 \approx I_3 \approx 1/2$. 
In contrast, if the coupling is sufficiently strong such that oscillators lock their 
phases, the actions ``unlock'' 
and chaos manifests itself in 
intermittent bursts of the actions $I_j(t)$ 
(cf.~Fig.~\ref{fig:3osc2} (b), (d)). 
For intermediate coupling strengths, regular regions still exist around 
$\phi_2 - \phi_1 \approx 3\pi/2$ (indicated by the trajectories colored 
in blue and green in Fig.~\ref{fig:phase} (b)) which confine the chaotic 
region around the torus $T^N$ and lead to episodes of seemingly
regular dynamics between the bursts.

\begin{figure}[tb]
\centering
\includegraphics[width=8cm, angle=0]{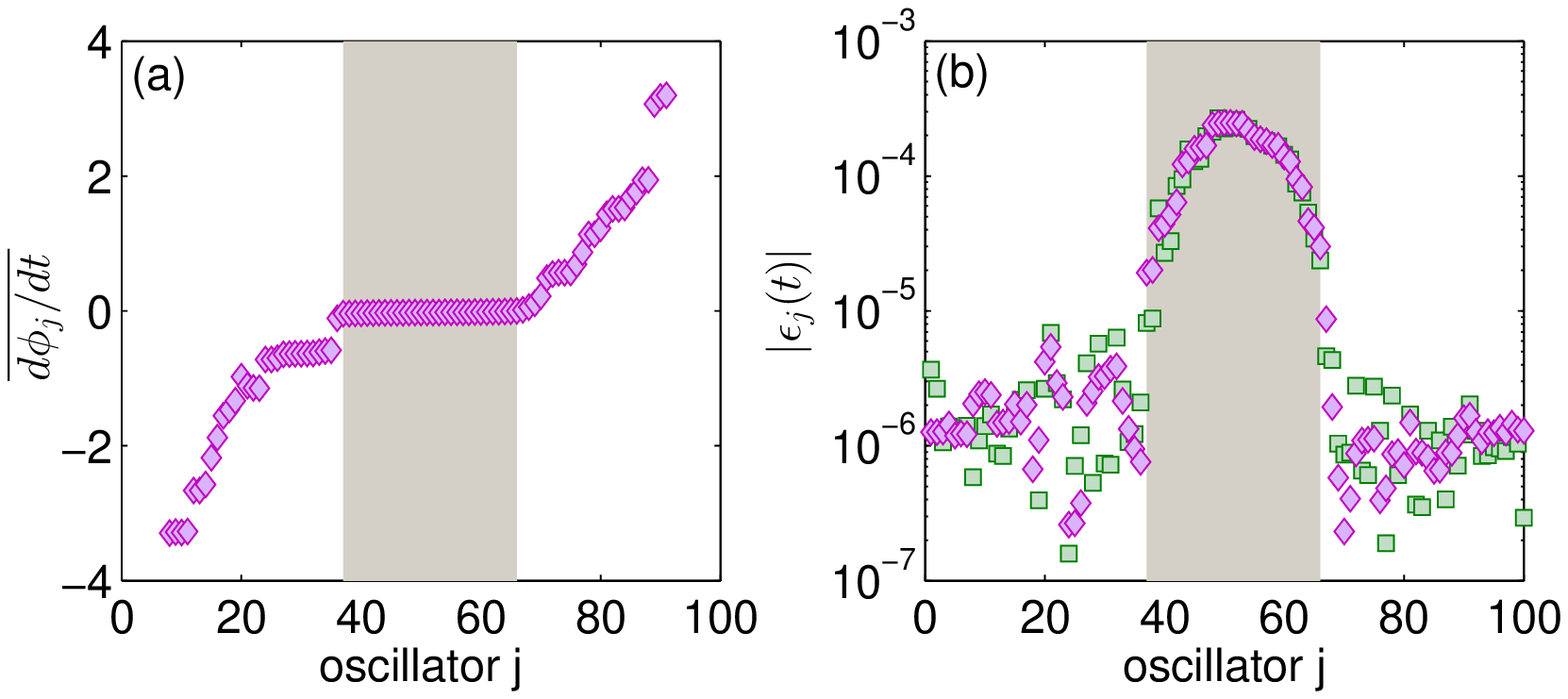}
\includegraphics[width=7cm, angle=0]{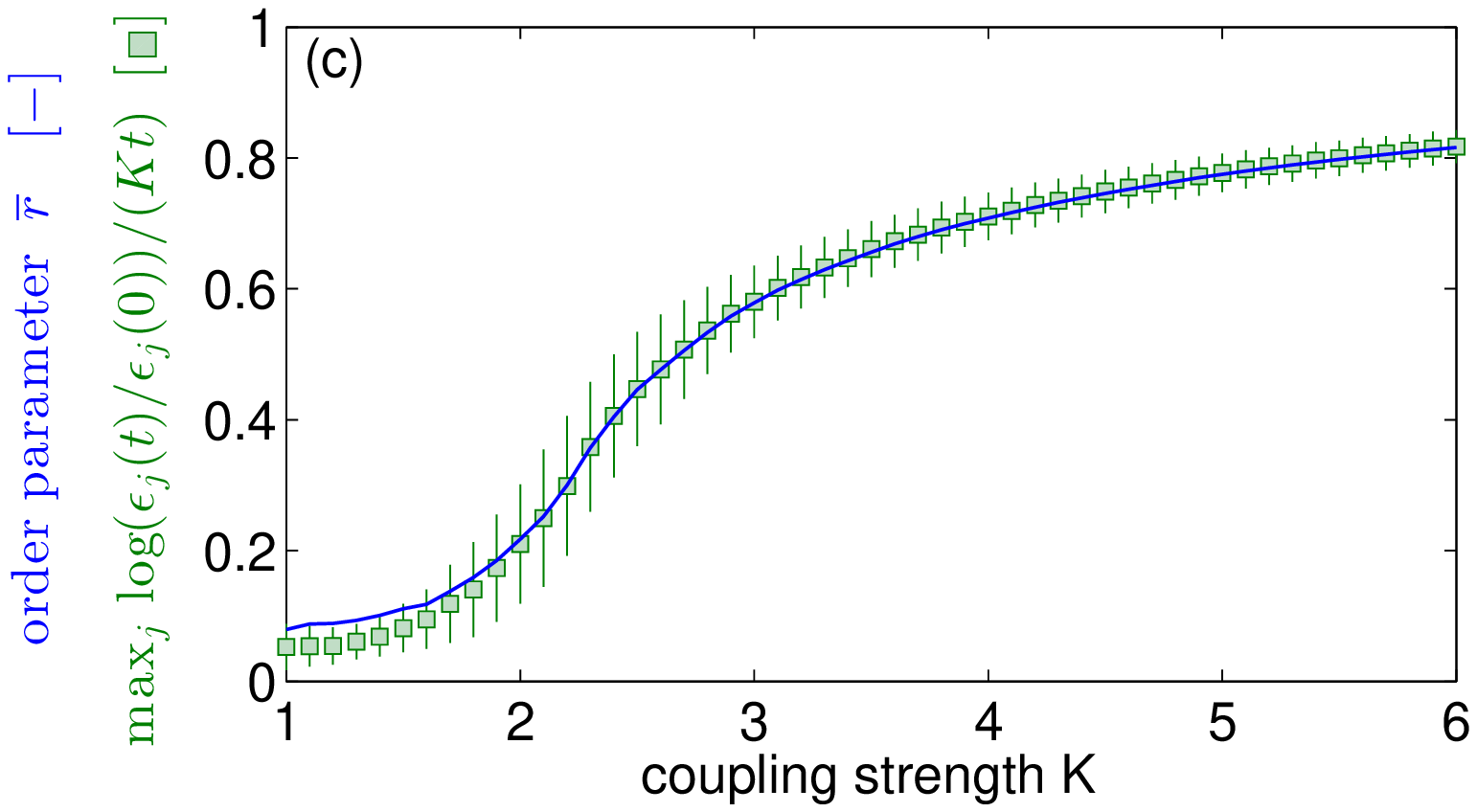}
\caption{\label{fig:predict}
Transverse instability of actions predicts order parameter.
(a) Average phase velocity $\overline{d \phi_j/dt}$ of 
$N=100$ oscillators in the regime of partial phase locking at
$K=2.2$. 
(b) Perturbations $|\epsilon_j(t)|$ grow off
the Kuramoto manifold, shown after $t=10$, starting from 
$\epsilon_j(0) = \pm 10^{-6}$ with random signs.
 Exact numerical results for the Hamiltonian
dynamics ($\square$) are displayed in comparison 
to the diagonal approximation 
(\ref{eqn:eom-eps2},\ref{eqn:Adiag}), ($\diamond$). 
Actions grow substantially more for those oscillators 
that are phase-locked (grey area).
(c) The prediction of the order parameter from the Hamiltonian action
dynamics (\ref{eqn:r0reconstruction}) ($\square$) well agrees with the actual order parameter $r$ (\ref{eq:standardOP}) directly measured from the Kuramoto phases. ($N=250$; data averaged over 100 realizations of the $\omega_j$; 
vertical lines indicate standard deviation of (\ref{eqn:r0reconstruction})) \cite{SimPar}.
}
\end{figure}

\begin{figure}[tb]
\centering
\includegraphics[width=8cm, angle=0]{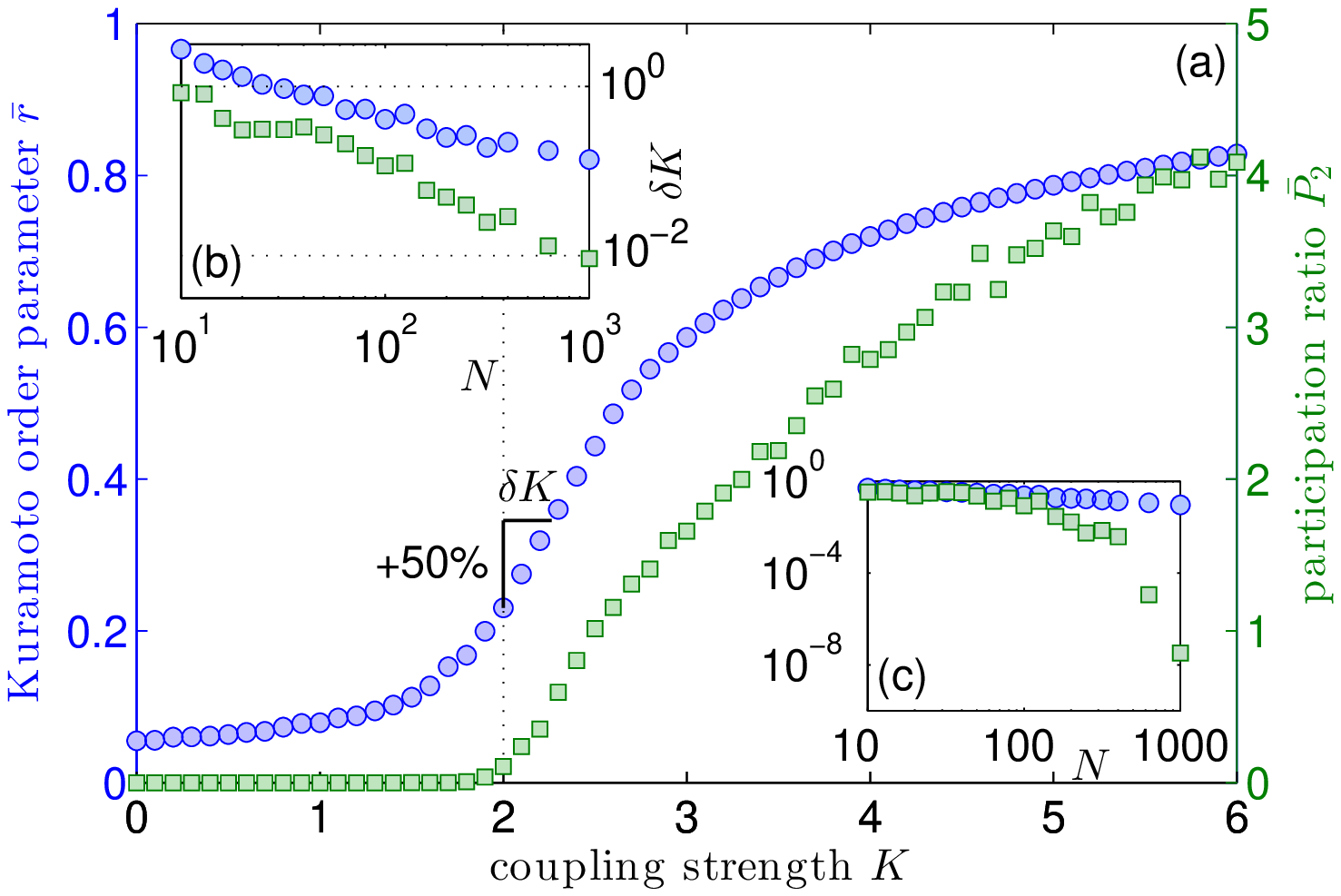}
\caption{\label{fig:khorder}
(Color online)
Participation ratio of Hamiltonian dynamics reveals synchronization phase transition. 
(a) Order parameter $\overline r$ ($\circ$, left scale) and  
participation ratio $\overline \Ps_2$ ($\square$, right scale) as a
function of coupling strength $K$ for $N = 250$ oscillators. 
Mean-field theory predicts the onset of phase 
order in the Kuramoto model at $K_c=2$ (dashed vertical line). 
Inset (b): width of the transition region $\delta K$, where
 $\overline r$ increases by 50 \% 
from its critical value at $K_c$, i.e.~$\overline r(K_c + \delta K) = 1.5 \, \overline r(K_c)$ 
as a function of the number of oscillators $N$. 
Inset (c): Finite size scaling of $\overline r$ and $\overline \Ps_2$ for 
a subcritical coupling strength $K = 1.8$.
Initial actions are $I_j = 1/2$ ($\circ$) and 
$I_j = 1/2 + \epsilon_j$ ($\square$), where the $\epsilon_j$ 
are drawn from a Gaussian distribution with standard 
deviation $10^{-4}$.
All data points are averages over 100 realizations each \cite{SimPar}.
}
\end{figure}

Bursts in the $I_j$ dynamics close to the
Kuramoto torus indicate phase
locking on the Kuramoto manifold also for general $N$ (not shown). 
As the full Hamiltonian flow conserves the phase space volume and the 
dynamics on $T^N$ is contracting around the phase locked state, it 
must be expanding in directions transverse to $T^N$. These expansions
are observed as bursts of the $I_j$.

To quantitatively understand the relation between action bursting and 
synchronization of the phases, we analytically derive the approximate dynamics of perturbations 
$$
   \epsilon_j(t) = I_j(t)-I_0
$$ 
off the torus $T^N$, where $I_0=1/2$ via action normalization. Expanding the equations of motion (\ref{eqn:eom_iphi}) to 
first order in $\epsilon_j$ (around $\epsilon_j\equiv 0$) yields 
the dynamics
\bea
       \label{eqn:eom-eps2}
    \dot \epsilon_j &=& \sum \nolimits_\ell \, A_{j,\ell}(t) \epsilon_\ell \; ,
\eea
where
  $  A_{j,\ell}(t) := \frac{K}{N} \left(  
              \delta_{j,\ell} \left[ \sum_m \cos(\phi_j - \phi_m) \right]
                           - \cos(\phi_j - \phi_\ell) \right).  \nn
$
This expression directly links the action instability to the phase locking 
dynamics of the original Kuramoto model: For small perturbations 
$\epsilon_j$, we approximate the dynamics of the phases $\phi_j(t)$ by the 
associated dynamics $\phi_j^T(t)$ on the Kuramoto manifold 
$T^N$. 
Assuming that the phase dynamics is fast, we further approximate
$A_{j,\ell}(t)$ by its time-average
\be
   \overline A_{j,\ell} := 
    \frac{K}{N} \; \overline{
              \delta_{j,\ell} \left[ \sum_m  \cos(\phi_j^T - \phi_m^T) \right]
                           - \frac{K}{N}\cos(\phi_j^T - \phi_\ell^T)}
   \label{eqn:Abar}
\ee

The structure of the matrix $\overline A_{j,\ell}$ becomes particularly 
simple if $N$ becomes large. The off-diagonal
elements decay as $1/N$ such that the matrix tends to be diagonal.
Carrying out the sum in the diagonal terms yields
\be
   \overline A_{j,\ell} = \delta_{j,\ell} \, K \; 
    \overline{ r \cos(\phi_j^T - \psi) }
    +\mathcal{O}\left( \tfrac{1}{N} \right),
  \label{eqn:Adiag}
\ee
with order parameter
\be
r e^{i \psi} = \frac{1}{N} \sum \limits_{m=1}^N  \, e^{ i \phi_m^T}
\label{eq:standardOP}
\ee
 as in  \cite{Stro00,Aceb05}.

In general, if the phase of an oscillator $j$ is not locked to the
overall phase $\psi$, the cosine tends to average out (with time) 
such that $\overline A_{j,j} \approx 0$.  On the
contrary, $\overline A_{j,j} > 0$ if the oscillator $j$ is locked.
Therefore we find that the perturbation $\epsilon_j(t)$ grow
exponentially if and only if the corresponding phases are locked, at least for $N\gg 1$.

The numerical example shown in 
Fig.~\ref{fig:predict} (a,b) illustrates this reasoning: Perturbations are particularly large, where  the associated Kuramoto oscillators are phase-locked (shaded regions). In particular, the fastest rate of divergence of the actions from the invariant manifold is expected for those oscillators that (i) are locked and (ii) are closest to the overall phase $\psi$ (center of locking region) such that the cosine term is maximal. 

The largest eigenvalue $\lambda_1$ of the matrix 
$\left( A_{j,\ell} \right)_{j,\ell} $ dominates the rate of 
divergence of the actions for randomly chosen initial 
conditions close to the Kuramoto manifold. 
In the diagonal approximation (\ref{eqn:Adiag}) for 
$N\gg 1$ and assuming independence (on average) of  $\overline{ r }$ and 
$\overline{\cos(\phi_j^T - \psi) }$ we obtain
\be
  \overline{r} \approx   \frac{1}{K} \max_j\ { \overline{A}_{jj}  }
  \label{eqn:r0prediction}
\ee 
because $\phi_j^T - \psi \approx 0$ for the $j$ yielding the 
maximum and thus $\overline{ \cos(\phi_j^T - \psi)  } \approx 1$.
This expression explicitly maps the stability properties of the actions 
to the locking properties of the phases. Thus the growth of the action perturbations in the full Hamiltonian system predicts the synchronization order parameter via 
\be
  r \approx   \frac{1}{K t} \max_j \, \log [ \epsilon_j(t) / \epsilon_j(0) ]. 
  \label{eqn:r0reconstruction}
\ee 
Direct observation of the order parameter from the Kuramoto phases
shows excellent agreement (cf.~Fig.~\ref{fig:predict}c) with this prediction.

How can the instability be quantified beyond the linear approximation (\ref{eqn:eom-eps2})? 
Again, we compare the dynamics on the Kuramoto manifold $T^N$  (with initial actions $I_j(0) = 1/2$) with trajectories started in its immediate 
proximity  (initial actions $I_j(0) = 1/2 + \epsilon_j$), and measure how
much these dynamics deviate from each other by evaluating 
the variance of $(2I_j)$. For this quantity, the
inverse participation ratio, we find 
\be
    \Ps_2 := \langle (2 I_j)^2 \rangle_j - \langle 2 I_j \rangle_j^2 
               = \frac{1}{N} \sum \nolimits_j (p_j^2 + q_j^2)^2 - 1 
          \label{eqn:ipr}
\ee
due to action normalization. By construction, $\Ps_2=0$ on the torus $T^N$.   
$\Ps_2 > 0$ indicates that the trajectory leaves the torus $T^N$
and starts to burst. It is known that systems of Kuramoto oscillators exhibit
a phase transition from an incoherent to a synchronized state at some $K_c$ in the thermodynamic limit $N \rightarrow \infty$ \cite{Stro00}. For
finite $N$, however, the transition is strongly blurred and the order
parameter $r$ increases smoothly with $K$ (see also
Fig.~\ref{fig:khorder}). Strikingly, the same transition in the full Hamiltonian
system is substantially clearer as indicated by a
sharp increase of the participation ratio $\Ps_2 > 0$ from originally small values close to zero (Fig. \ref{fig:khorder}).
In fact, $\Ps_2$ indicates the transition more precisely than the Kuramoto order 
parameter, both with respect to the finite-size scaling below 
the transition and the jump occurring at the transition point $K_c$
(see the insets Fig.~\ref{fig:khorder}b,c).

In summary, we have analytically and numerically demonstrated that a family of  Hamiltonian systems bears the celebrated Kuramoto model of coupled oscillators as dynamics on its invariant (toric) manifolds $T^N_I$. Interestingly, the emergence of synchrony is equivalent to the emergence of a transverse instability off such a torus. Therefore, the divergence of the actions due to a transverse instability in the Hamiltonian system quantifies the synchronization order parameter of the exact Kuramoto dynamics on  $T^N_I$. Moreover, for finite systems the participation ratio provides a distinguished  indicator for the onset of transverse instability in the full Hamiltonian system and, consequently, the onset of 
 synchronization of the Kuramoto phase dynamics. 
Taken together, these results establish an exact relation between dissipative
Kuramoto systems and volume-preserving Hamiltonian systems and may thereby further our understanding of the intriguing properties of networks of phase oscillators.


\begin{thebibliography}{1}
\bibitem{psection}
$I_2$ is plotted vs. $\phi_2 - \phi_1$ whenever the 
phase $\phi_3 - \phi_1$ crosses zero with $\dot \phi_3 - \dot \phi_1> 0$.
$I_1$ and $I_3$ are then fixed due to the conservation of 
total action $C^2$ and the total energy $\HH$.
\bibitem{SimPar}
In the simulations, we record $\overline r$, $\overline \Ps_2$, $\overline A_{j,\ell}$ 
and $\overline{d \phi_j/dt}$ averaged over a time interval $\Delta t_2 = 10$, 
after an initial transient period of length $\Delta t_1 = 10$.
The initial $\phi_j$ are randomly drawn from the uniform distribution 
on $[0,2\pi)$, but the same for perturbed and unperturbed actions.
In Fig.~\ref{fig:predict} (b), we use phases $\phi_j^T$ after the transient 
as initial values.
In Fig.~\ref{fig:predict} (c), we integrate the Hamiltonian dynamics up 
to a time $t$ for which $\max_j |\epsilon_j(t)| \ge 10^{-4}$.
The frequencies $\omega_j$ are drawn from a Cauchy distribution 
with density $g(\omega) = (a/\pi)/(\omega^2 + a^2)$ and width $a=1$.
\end{thebibliography}
\end{document}